\documentclass[twocolumn,superscriptaddress,floatfix,preprintnumbers, nofootinbib,hyperref]{revtex4} 
\pdfoutput=1
\usepackage[colorlinks=true,breaklinks=true]{hyperref}
\usepackage[utf8]{inputenc}
\hypersetup{allcolors=[rgb]{0.0 0.0 0.6},linkcolor=[rgb]{0.75 0.05 0.05}}
\usepackage{amsmath,amssymb,amsfonts}
\usepackage{epsfig}  
\usepackage{float}
\usepackage{graphicx}   
\usepackage{slashed}             
\usepackage{url}
\usepackage{color}
\usepackage{multirow}
\usepackage[normalem]{ulem}
\usepackage[dvipsnames]{xcolor}
\usepackage{letltxmacro}
\LetLtxMacro{\oldcite}{\cite}
\renewcommand{\cite}[1]{\mbox{\oldcite{#1}}}

\newcommand{\beq}{\begin{equation}}
\newcommand{\eeq}{\end{equation}}

\allowdisplaybreaks

\bibpunct{[}{]}{,}{n}{}{,}

\begin{document}

\title{Searching for the radiative decay of the cosmic neutrino background
with line-intensity mapping
}

\author{Jos\'e Luis Bernal}\email{jbernal2@jhu.edu}
\affiliation{Department of Physics and Astronomy, Johns Hopkins University, 3400 North Charles Street, Baltimore, Maryland 21218, USA}

\author{Andrea Caputo}\email{andrea.caputo@uv.es}
\affiliation{School  of  Physics  and  Astronomy,  Tel-Aviv  University,  Tel-Aviv  69978,  Israel}
\affiliation{Department  of  Particle  Physics  and  Astrophysics,Weizmann  Institute  of  Science,  Rehovot  7610001,  Israel}%
\affiliation{Max-Planck-Institut f\"ur Physik (Werner-Heisenberg-Institut), F\"ohringer Ring 6, 80805 M\"unchen, Germany}

\author{Francisco Villaescusa-Navarro}
\affiliation{Department of Astrophysical Sciences, Princeton University, Peyton Hall, Princeton, NJ, 08544, USA}
\affiliation{Center for Computational Astrophysics, 162 5th Avenue, New York, NY, 10010, USA}
\author{Marc Kamionkowski}
\affiliation{Department of Physics and Astronomy, Johns Hopkins University, 3400 North Charles Street, Baltimore, Maryland 21218, USA}

\date{\today}
\smallskip


\begin{abstract}

We study the possibility to use line-intensity mapping (LIM) to seek photons from the radiative decay of neutrinos in the cosmic neutrino background.  The Standard Model prediction for the rate for these decays is extremely small, but it can be enhanced if new physics increases the neutrino electromagnetic moments.  The decay photons will appear as an interloper of astrophysical spectral lines.  We propose that the neutrino-decay line can be identified with anisotropies in LIM clustering and also with the voxel intensity distribution. Ongoing and future LIM experiments will have---depending on the neutrino hierarchy, transition and experiment considered---a sensitivity to an effective electromagnetic transition moment $\sim 10^{-12}\, -\,10^{-8}\, (m_ic^2/{0.1 \rm eV})^{3/2}\mu_{\rm B}$, where $m_i$ is the mass of the decaying neutrino and $\mu_{\rm B}$ is the Bohr magneton. This will be significantly more sensitive than cosmic microwave background spectral distortions, and it will be competitive with stellar cooling studies.  As a byproduct, we also report an analytic form of the one-point probability distribution function for neutrino-density fluctuations, obtained from the \textsc{Quijote} simulations using symbolic regression.

\end{abstract}

\maketitle


Considerable efforts are underway to study the properties of neutrinos, including their masses, mixing angles, and nature (e.g., Dirac or Majorana)~\cite{Yoshi75,Rosenberg:1962pp,Mertens:2018vuu, KamLAND-Zen:2016pfg, Albert:2014awa, Brunner:2017iql,Beda:2009kx, Deniz:2009mu, Abe:2017vif, Adamson:2017gxd, Acciarri:2016crz, Abe:2015zbg, Aartsen:2017bap, Raffelt:1996wa, Pontecorvo:1967fh, Gribov:1968kq, Langacker:1986jv, Bilenky:1980cx, Akhmedov:1999uz, GonzalezGarcia:2002dz, GonzalezGarcia:2007ib}.  The stability of neutrinos is also of interest.  
An active massive neutrino $\nu_i$ can decay into a lighter eigenstate $\nu_j$ and  photon, $\gamma$, $\nu_i\rightarrow \nu_j+\gamma$ with a rate determined by electromagnetic transition moments induced via loops involving gauge bosons. 
The Standard Model (SM) prediction for 
the lifetime is $\tau_{\rm SM} = 7.1\times 10^{43}m^{-5}_{\rm eV}$ s~\cite{1992pmn..book.....B, Pal:1981rm, Petcov:1976ff, Fujikawa:1980yx}, where $m_{\rm eV}\equiv m_\nu c^2/{\rm eV}$ is the neutrino mass in eV$/c^2$ units, significantly longer than the age of the Universe.

However, new physics beyond the SM (BSM) can enhance neutrino magnetic moments~\cite{Bell:2006wi,Davidson:2005cs,Georgi:1990se,Lee:1977tib,Shrock:1974nd,Lindner:2017uvt,Lindner_2017,Miranda:2020kwy,Babu:2020ivd} 
and such modifications have been considered in connection with experimental anomalies, such a possible correlation of solar neutrinos with Solar activity~\cite{Bahcall:1989ks, Raffelt:1996wa}, or more recently~\cite{Miranda:2020kwy,Babu:2020ivd} the $\sim 3\sigma$ excess reported XENON1T~\cite{Aprile:2020tmw}.  Although many avenues have been proposed (see e.g., Ref.~\cite{Giunti:2014ixa} for a review), the most efficient direct laboratory probe of neutrino electromagnetic couplings involves neutrino-electron scattering~\cite{PhysRevD.39.3378, Beda:2009kx, Deniz:2009mu}.  Tighter bounds on neutrino electromagnetic moments come from astrophysics. In particular, the strongest constraint comes from the tip of the red giant branch in globular clusters, which is sensitive to the additional energy loss through plasmon decay into two neutrinos~\cite{PhysRevLett.64.2856,1992A&A...264..536R,Viaux:2013hca}. Radiative neutrino decays have also been constrained from measurements of the cosmic microwave background (CMB) spectral distortions~\cite{Mirizzi:2007jd, Aalberts:2018obr}. 

Here we study the use of line-intensity mapping (LIM) to seek photons from radiative decays of neutrinos in the cosmic neutrino background.  
LIM~\cite{Kovetz:2017agg, Kovetz:2019uss} exploits the integrated intensity at a given frequency induced by a well-identified spectral line to map the three-dimensional distribution of matter in the Universe.  Photons from particle decays will appear in these maps as an unidentified line~\cite{Creque-Sarbinowski_LIMDM} that can be distinguished from astrophysical lines through its clustering anisotropies and through the voxel probability distribution function~\cite{Bernal_LIMDM}.  We find that LIM has the potential to be significantly more sensitive to radiative decays than current cosmological probes and compete with the strongest bounds to electromagnetic moments coming from astrophysical observations.

While neutrino radiative decays are characterized by the electromagnetic transition moments, LIM experiments are sensitive to the luminosity density $\rho_{\rm L}$ of the photons produced in each point $\boldsymbol{x}$, which, for the decay between the $i$ and $j$ states, is given by
\begin{equation}
    \rho_{\rm L}^{ij}(\boldsymbol{x}) = (1/6)\rho_{\nu}(\boldsymbol{x})c^2 
    \Gamma_{ij}\left(1-m_j^2/m_i^2\right)\,,
    \label{eq:rhoLnu}
\end{equation}
where $\rho_{\nu}$ is the total neutrino density, $\Gamma_{ij}\equiv\tau_{ij}^{-1}$ is the decay rate, and $m_i$ are the neutrino masses. We assume that the density of each state is $1/3$ of the total density, as expected apart from small mass differences and flavor corrections that have negligible consequences for the precision goals of this \textit{Letter}~\cite{Mangano:2006ar}. The corresponding brightness temperature $T$ at redshift $z$ is
\begin{equation}
\begin{split}
     T^{ij}(z,\boldsymbol{x}) & =  \frac{c^3(1+z)^2\rho_{\rm L}^{ij}(z,\boldsymbol{x})}{8\pi k_{\rm B}f^3H(z)} = X_{\rm LT}\rho^{ij}_{\rm L}(z,\boldsymbol{x}) = \\
     & = %
     (X_{\rm LT}/6)\rho_{\nu}c^2 
    \Gamma_{ij}\left(1-m_j^2/m_i^2\right)\,,
\end{split}
\label{eq:T}
\end{equation}
where $H$ is the Hubble expansion and $k_{\rm B}$ is the Boltzmann constant and $f$ is the rest-frame frequency~\cite{Lidz_2011}. Thus, the brightness temperature from neutrino decays traces the neutrino density field. 

\begin{figure}[t]
\centering
  \includegraphics[width=0.95\linewidth]{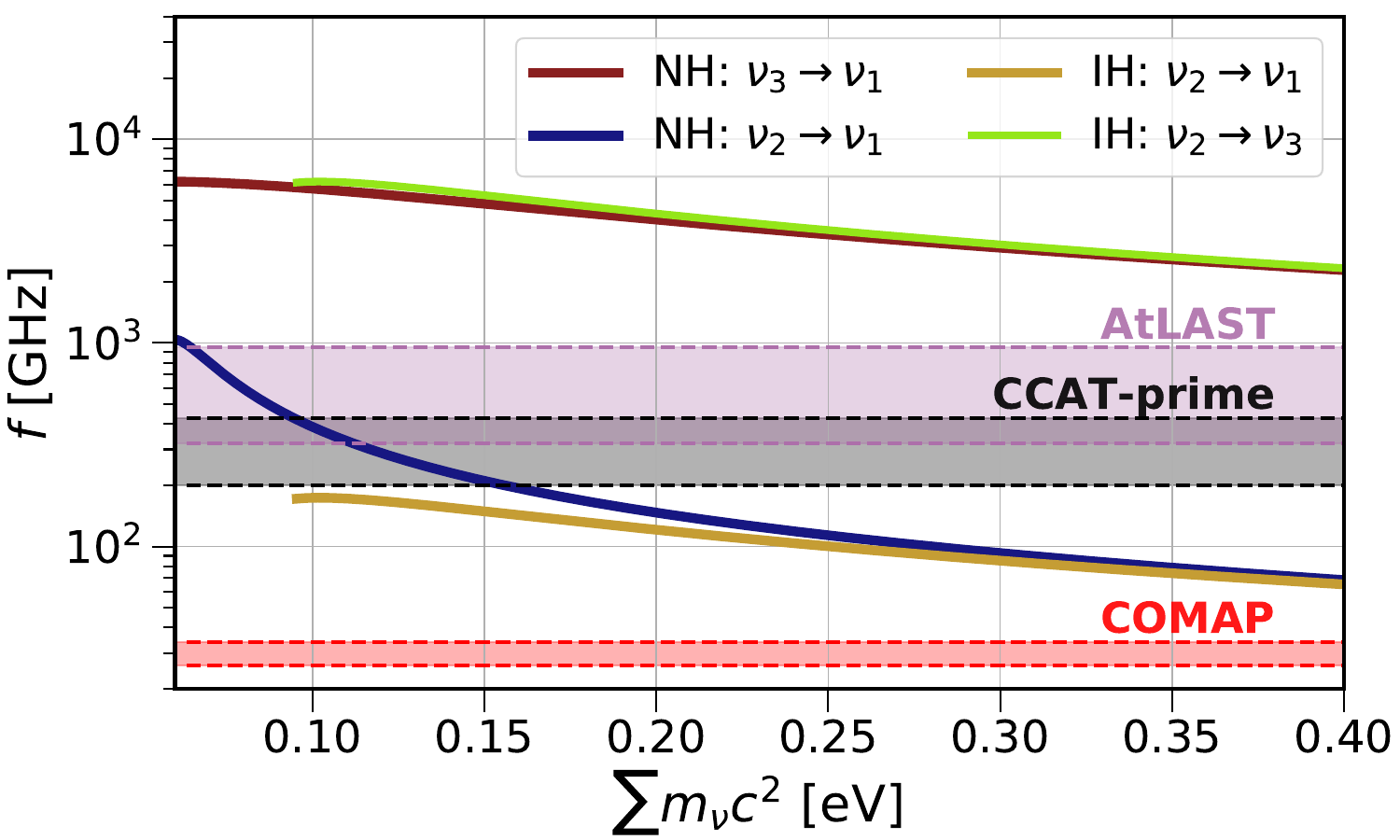}
  \includegraphics[width=0.95\linewidth]{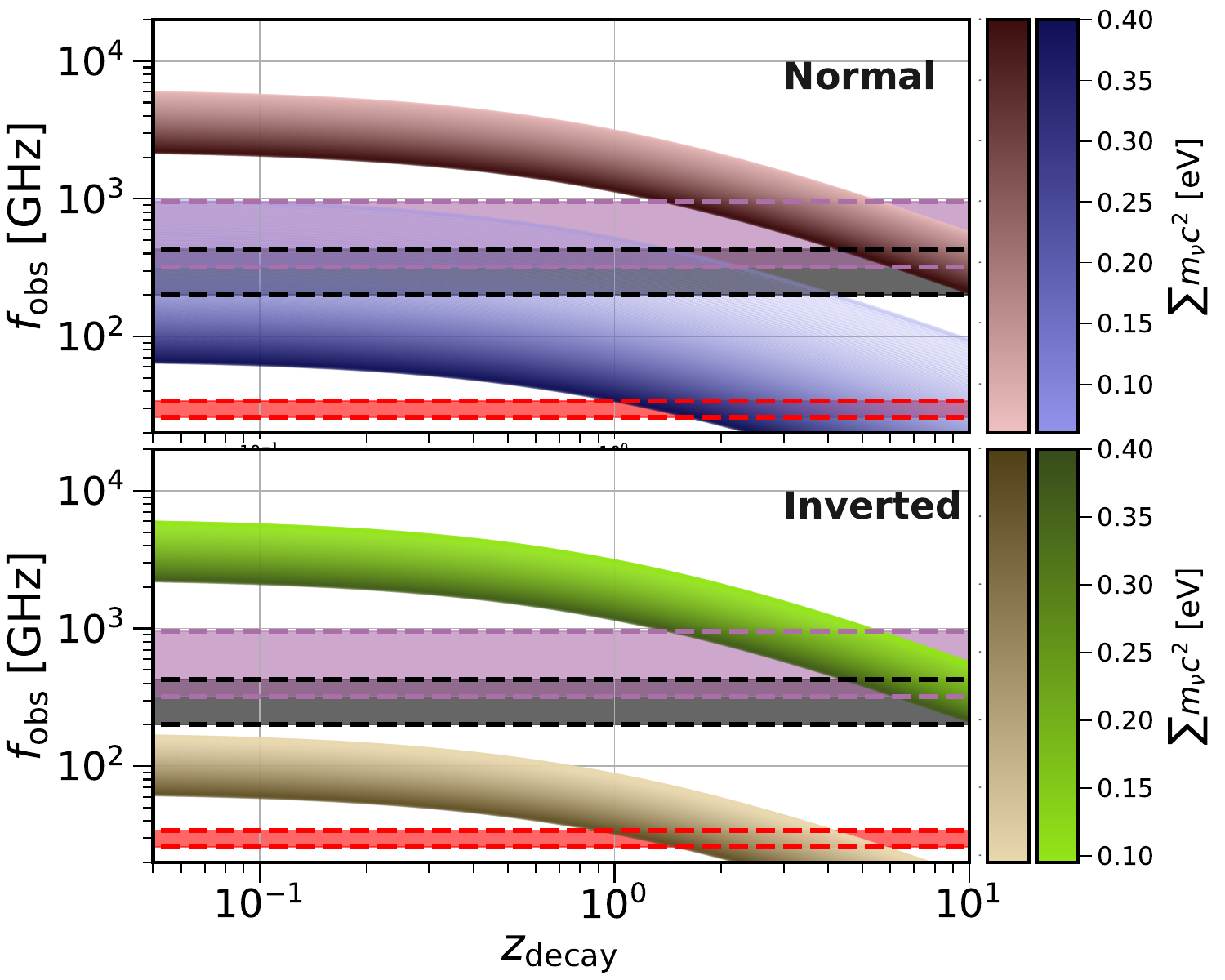}
  \vspace{-0.3cm}
\caption{Relation between the sum of neutrino masses and the rest-frame frequency of the photon produced in the decay for normal (NH) and inverted (IH) hierarchies (top) and the corresponding observed frequency. We also show the frequency bands of the experiments considered in both panels with horizontal shaded bands.}
\label{fig:mtotnu}
\end{figure}

Decay photons are then an emission line with rest-frame frequency given by $f_{ij}=\left(m_i^2-m_j^2\right)c^2/\left(2h_{\rm P}m_i\right)$, where $h_{\rm P}$ is the Planck constant. For $m_i/c^2\gg T_{\nu}/k_{\rm B} \sim 10^{-4}(1+z)\,{\rm eV}$ (where $T_\nu$ is the cosmic neutrino temperature), which holds true for our cases of interest, the neutrinos are non-relativistic and we can neglect the linewidths due to their velocity dispersion.\footnote{The  widening would only be relevant if larger than the instrumental spectral resolution $f_{\rm obs}/\delta f$, where $f_{\rm obs}$ is the observed frequency and $\delta f$ is the channel width. In such case, we would take the line width as the spectral resolution for the neutrino decay line.} The rest-frame frequency of the emission lines is then uniquely characterized by the neutrino hierarchy and the sum $\sum m_\nu$ of neutrino masses, as shown in Fig.~\ref{fig:mtotnu}, with the observed frequency redshifted accordingly. The transitions not included in the figure have a very similar frequency than one of the other two (e.g., $f_{31}\approx f_{32}$ for the normal hierarchy) and are not distinguished hereinafter. 

We now consider two LIM observables: the power spectrum and the voxel intensity distribution (VID). The observed anisotropic LIM power spectrum associated to the neutrino decay between $i$ and $j$ states is~\cite{Bernal_IM, Bernal_LIMDM}
\begin{equation}
    P_{ij}(k,\mu) = W(k,\mu)X_{\rm LT}^2\langle \rho_{\rm L}^{ij}\rangle^2 F^2_{\rm rsd}(k,\mu)P_{\rm \nu}(k)\, ,
    \label{eq:DMPK}
\end{equation}
where $k$ is the modulus of the Fourier mode, $\mu \equiv \boldsymbol{k}\cdot\boldsymbol{k}_\parallel/k^2$ is the cosine of the angle between the Fourier mode and the line of sight, $W$ is a window function modeling the effects from instrumental resolution and finite volume observed, the brackets $\langle \rangle$ denote the spatial mean, $F_{\rm RSD}$ is a redshift-space distortions factor~\cite{Bernal_IM}, $P_{\rm \nu}$ is the neutrino power spectrum, computed using \texttt{CAMB}~\cite{Lewis:1999bs}, and all redshift dependence is implicit. We consider the Legendre multipoles of the LIM power spectrum with respect to $\mu$ up to the hexadecapole. 

\renewcommand{\arraystretch}{1.5}
\begin{table}[]
\centering
\resizebox{\columnwidth}{!}{%
\begin{tabular}{|l||c|c|c|}
\hline
Experiment & COMAP 1 (2) & CCAT-prime & AtLAST \\ \hline
Line & CO & CII & CII \\ \hline
Freq. band {[}GHz{]} & 24-36 & \begin{tabular}[c]{@{}c@{}}200-240, 260-300,\\ 330-370, 388-428\end{tabular} & \begin{tabular}[c]{@{}c@{}}315-376, 376-470,\\ 470-620, 620-920\end{tabular} \\ \hline
Spectral resolution & 4000 & 100 & 1000 \\ \hline
Ang. resolution {[}''{]} & 240 & 57, 45, 35, 30 & 4.4, 3.6, 2.8, 2.0 \\ \hline
Sky coverage {[}deg$^2${]} & 2.25 (60) & 8 & 7500 \\ \hline
Voxel noise & 39 (69) $\mu$K & \begin{tabular}[c]{@{}c@{}}(0.6, 1.0, 2.5, \\ 5.7)$\times10^4$ Jy/sr \end{tabular} & \begin{tabular}[c]{@{}c@{}}(0.4, 0.7, 1.4, \\ 3.9)$\times10^5$ Jy/sr\end{tabular}  \\ \hline
\end{tabular}%
}
\caption{Instrumental specifications used. Each independent frequency band is separated by commas. CII observations use the convention for specific intensity instead of brightness temperature. We combine 30$\times$30 pixels for VID analyses with AtLAST. More details can be found in Ref.~\cite{Bernal_LIMDM}.}
\label{tab:exp}
\end{table}
\renewcommand{\arraystretch}{1}

Similarly, the VID is related to the probability distribution function (PDF) $\mathcal{P}_{\breve{\rho}}$ of the normalized total neutrino density $\breve{\rho}_{\rm \nu}\equiv \rho_{\rm \nu}/\langle{\rho}_{\rm \nu}\rangle$, as $\mathcal{P}_{ij}(T) = \mathcal{P}_{\breve{\rho}}(\breve{\rho}_{\rm \nu})/\langle T^{ij}\rangle$. We estimate the neutrino density PDF from high-resolution simulations of the \textsc{Quijote} simulation suite~\cite{Villaescusa-Navarro:2019bje}, that model the gravitational evolution of more than 2 billion cold dark matter and neutrino particles in a comoving box of $(1~h^{-1}{\rm Gpc})^3$ volume.  Degenerate neutrino mass eigenstates are assumed. 

First, neutrino particle positions are assigned to a regular grid with $1500^3$ voxels employing the cloud-in-cell mass-assignment scheme. Next, the 3D field is convolved with a Gaussian kernel of a given width. Then, the PDF is estimated by computing the fraction of voxels with a given $\breve{\rho}_\nu$. We do this for $\sum m_\nu c^2 = \lbrace 0.1,\, 0.2,\, 0.\,4\rbrace\, {\rm eV}$, at $z = \lbrace 0,\, 0.5,\, 1,\, 2,\, 3,\, 4,\, 5,\, 6,\, 7,\, 8,\, 9\rbrace$ and for 6 smoothing scales $\lbrace2,\, 3,\, 4,\, 5,\, 7.5,\, 10\rbrace\, h^{-1}{\rm Mpc}$. We have checked that the computed PDFs, in the range of interest for this study, are converged in our simulations. Note that all  dependences can be condensed in the root-mean square $\sigma$ of smoothed density field, which depends on $\sum m_\nu$, $z$ and the smoothing scale. Finally, we use symbolic regression to approximate this grid of PDFs using the Eureqa package (\url{https://www.datarobot.com/nutonian/}) finding
\begin{eqnarray}
     \frac{\mathcal{P}_{\breve{\rho}}}{\mathcal{A}}&  =& \exp\left\lbrace \frac{0.2\mathcal{G}(0.6\frac{\mathfrak{d}}{\mathfrak{s}})+2.5\mathfrak{s}^{1.6}\mathcal{G}(1.1+\frac{\mathfrak{d}}{\mathfrak{s}}-2.3\mathfrak{s})}{\mathfrak{s}+0.05\mathcal{G}(0.6\frac{\mathfrak{d}}{\mathfrak{s}})} \right. \nonumber \\
    & &  -2.5\mathfrak{s}^{1.6}\mathcal{G}\left(1.1+(\mathfrak{d}/\mathfrak{s})-2.3\mathfrak{s}\right) \bigg\}-1,
\end{eqnarray}
where $\mathcal{G}(x) \equiv e^{-x^2}$, $\mathfrak{d}\equiv \log \breve{\rho}_\nu $, $\mathfrak{s}\equiv \log (1+\sigma)$, and $\mathcal{A}$ is a normalization factor.

LIM experiments will not target the emission line from neutrino decays, but known astrophysical lines. In turn, the neutrino decay line will redshift into the telescope frequency band from a different redshift. All  emission lines other than the main target that contribute to the total signal tracing other cosmic volumes are known as line interlopers. These contributions, if known, can be identified and modeled (see e.g.,~\cite{Oh2003,Wang2006,Liu2011,Breysse_foregrounds, Lidz_interlopers, Sun_foregrounds, Cheng_foregroundsAP, Cheng_deconfInterlopers, Gong_interlopers}). However, the neutrino decay line will be an unknown line interloper. From Fig.~\ref{fig:mtotnu} we can see that the frequencies of interest lie in the frequency bands of experiments like COMAP~\cite{Cleary_COMAP} (which targets the CO line) and CCAT-prime~\cite{CCAT-prime} and AtLAST~\cite{ATLAST} (which target the CII line); their instrumental specifications are summarized in Table~\ref{tab:exp}. 

We assume the fiducial astrophysical model for the CO and the CII lines from Refs.~\cite{Li_CO_16} and~\cite{Silva_CII}, and model their power spectrum and VID, with their corresponding covariances, following Refs.~\cite{Bernal_IM,Breysse_VID, Bernal_LIMDM}. For the VID analysis, we use a modified Schechter function with the parameters reported in Ref.~\cite{Bernal_LIMDM}. We take $\Lambda$CDM cosmology with best-fit parameter values from \textit{Planck} temperature, polarization and lensing power spectra~\cite{Planck18_parameters} assuming $\sum m_{\nu}c^2=0.06$ as our fiducial model. We consider normal (NH) and inverted (IH) neutrino hierarchies.\footnote{In the Fisher-matrix analysis, the variation of $\sum m_{\nu}$ and the change of the neutrino hierarchy are included in our fiducial model: we only consider deviations due to the neutrino decay and not to the varying neutrino masses.} 

\begin{figure*}[t]
\hspace*{-1cm}
\centering
  \includegraphics[width=0.5\linewidth]{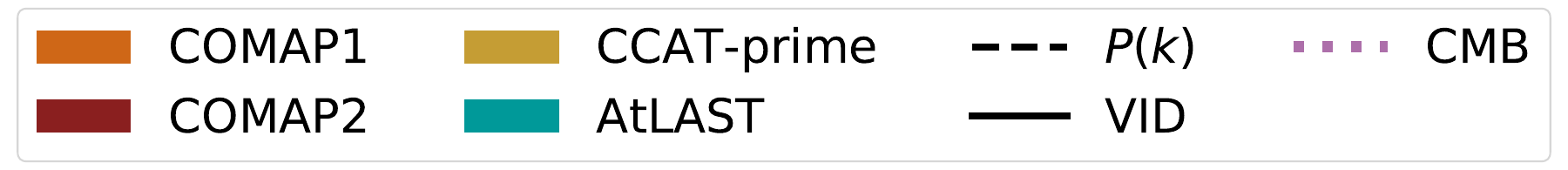}
  \includegraphics[width=0.85\linewidth]{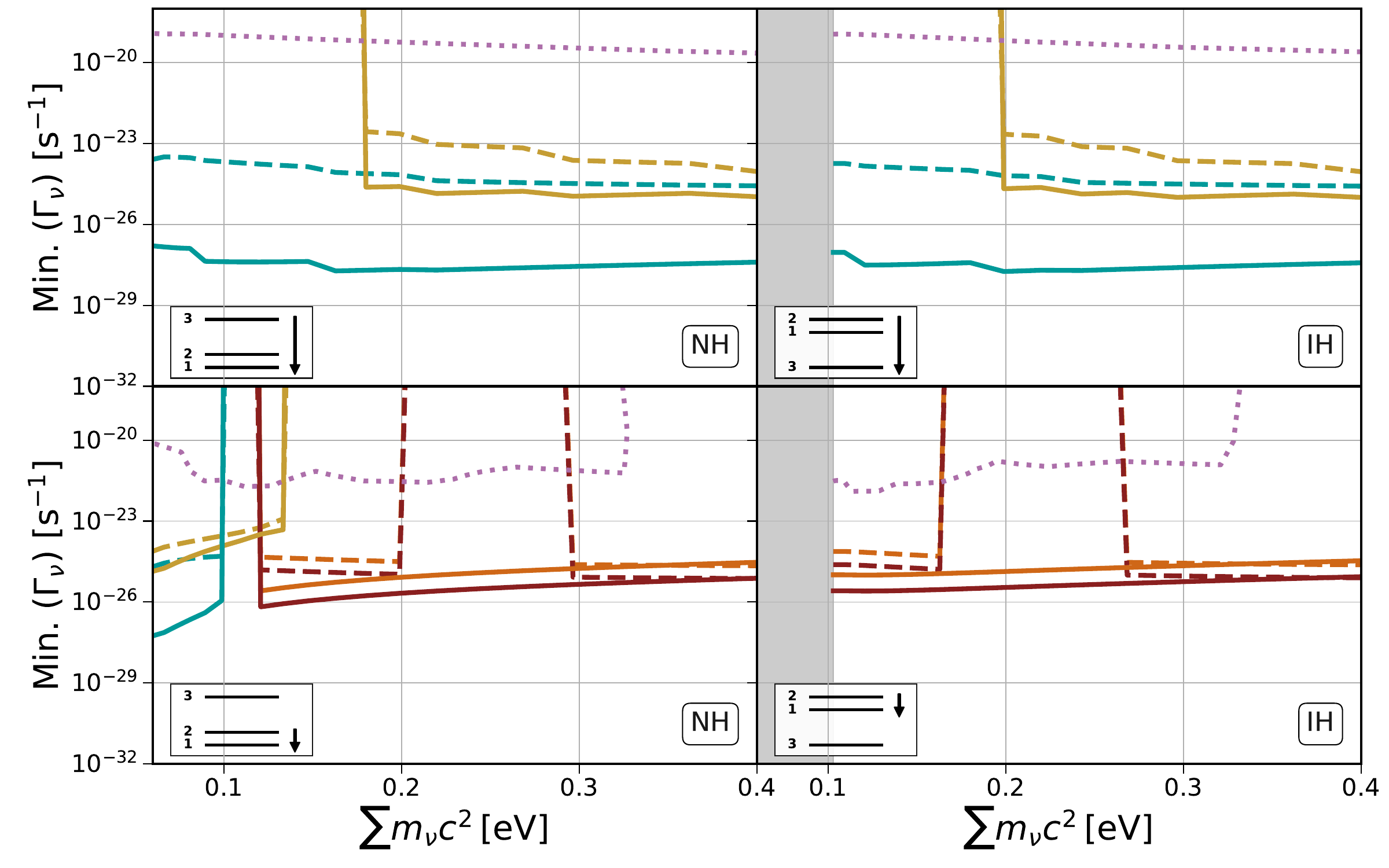}
  \vspace{-0.5cm}
\caption{\label{fig:sunproperties}
Forecasted 95\% confidence level marginalized upper limits of $\Theta_{\nu}$ as function of the total neutrino mass from measurements for the power spectrum (dashed lines) and VID (solid lines) for all LIM surveys considered, namesly COMAP1 (orange), COMAP2 (red), CCAT-prime (dark yellow) and AtLAST (light blue). The left panels refer to the NH case, while the right ones to the IH case. In both cases we also indicate the considered transition between eigenstates. The dotted purple lines indicate the CMB limits from ref.~\cite{Mirizzi:2007jd, Aalberts:2018obr}.}
\label{fig:results}
\end{figure*} 

Recently, a similar situation, regarding decaying dark matter, was described in Ref.~\cite{Bernal_LIMDM}, where strategies to detect such decays were proposed. Here we adapt that modeling to the neutrino decay case, considering neutrino decays happening at $z<10$, and perform a Fisher-matrix analysis~\cite{Fisher:1935, Jungman:1995av,Jungman:1995bz, Tegmark_fisher97}, accounting for the uncertainity in the astrophysical model.  In summary, the contribution from neutrino decays to the VID can be modeled by convoluting $\mathcal{P}_{ij}(T)$ with the astrophysical and noise VIDs: the total VID is the result of the sum of the three contributions. In turn, the contribution to the power spectrum consists of the addition of the projected power spectrum from neutrino decays to a different redshift, which introduces a strong anisotropy in the power spectrum, altering the ratio between the Legendre multipoles. For the power spectrum, we do not consider decays from the same cosmic volumes probed by the astrophysical line because they are very degenerate with astrophysical uncertainties.

We show the forecasted minimum values of $\Gamma_{ij}$  which LIM experiments will be sensitive to at the 95\% confidence level, as function of the neutrino hierarchy, transition, and $\sum m_\nu$ in Fig.~\ref{fig:results}. We limit the minimum $\sum m_\nu$ at the minimum mass allowed for each hierarchy from neutrino oscillations experiments~\cite{Esteban:2018azc}.  As expected from Fig.~\ref{fig:mtotnu}, COMAP and the experiments targeting CII are sensitive to the transitions between close and far mass eigenstates, respectively (with the exception of low $\sum m_{\nu}$ in the normal hierarchy). 

After marginalizing over the astrophysical uncertainties of the target line as in Ref.~\cite{Bernal_LIMDM}, we find that for all cases considered LIM experiments can improve current cosmological bounds on the neutrino decay rate from CMB spectral distortions~\cite{Mirizzi:2007jd, Aalberts:2018obr} by several orders of magnitude. This shows that LIM has the potential to provide the strongest cosmological sensitivity on neutrino radiative decays. Furthermore, LIM will be competitive to the most stringent limits to date, coming from stellar cooling~\cite{PhysRevLett.64.2856,1992A&A...264..536R,Viaux:2013hca}, as we see below.

As mentioned above, at the microscopic level radiative neutrino decays may result from an effective term in the Lagrangian like $\propto\bar{\nu}^i \sigma_{\alpha\beta}(\mu_{ij} + \epsilon_{ij}\gamma_5) \nu^j F^{\alpha\beta} +$ hermitian conjugates~\cite{Giunti:2014ixa, Raffelt:1996wa,Brdar:2020quo}, where $F^{\alpha\beta}$ is the electromagnetic field tensor, $\sigma_{\alpha\beta}$ is the Dirac gamma matrices commutator, and $\mu_{ij}$ and $\epsilon_{ij}$ are the magnetic and electric moments, respectively. For a transition (i.e., $i\neq j$), we can relate an effective electromagnetic moment $\mu^{\rm eff}_{ij}$ to the decay rate as
\begin{equation}
    \left( \mu^{\rm eff}_{ij} \right)^2 \simeq \frac{\Gamma_{ij}}{5\,{\rm s^{-1}}}\frac{m_{{\rm ev,}i}^3}{\left(m_{{\rm eV,}i}^2-m_{{\rm eV,}j}^2\right)^3}\mu_{\rm B}^2\,,
    \label{eq:Gamma-mu}
\end{equation}
where $\lvert\mu^{\rm eff}_{ij}\lvert^2\equiv \lvert\mu_{ij}\lvert^2+\lvert\epsilon_{ij}\lvert^2$, and $\mu_{\rm B}$ is the Bohr magneton.

According to Eq.~\eqref{eq:Gamma-mu}, the forecasted LIM sensitivity of $\Gamma_{ij}\sim 10^{-28}\,-\,10^{-25}\, {\rm s}^{-1}$ at 95\% confidence level translates to $\mu_{ij}^{\rm eff} \sim 10^{-12}\, -\,10^{-8}\, (m_ic^2/{0.1 \rm eV})^{3/2}\mu_{\rm B}$, while current and forecasted CMB limits are $\sim 10^{-7}\,-\,10^{-8} \mu_{\rm B}$ and $\sim 10^{-8}\,-3\times 10^{-11}\mu_{\rm B}$, respectively~\cite{Aalberts:2018obr}. Note that the sensitivity to $\mu^{\rm eff}_{ij}$ depends on the mass of the original neutrino, which in turn depends on the transition, hierarchy and $\sum m_\nu$ considered. In turn, the most stringent direct detection limit was obtained in the Borexino experiment and is related to an effective moment accounting for all magnetic direct and transition moments: $\mu^{\rm eff}_\nu<2.8\times 10^{-11}\mu_{\rm B}$ at 90\% confidence level~\cite{Borexino:2017fbd}. Finally, astrophysical studies of stellar cooling set the strongest bounds to date: $\mu^{\rm eff}_\nu< 4.5\times 10^{-12} \mu_{\rm B}$ at 95\% confidence level~\cite{Viaux:2013hca}. 

This demonstrates the great potential that LIM surveys have to unveil neutrino properties: on top of having a sensitivity competitive to and in some cases even improving current strongest limits, LIM experiments may probe neutrino decays in a very different context than the rest of experiments and observations discussed above. Instead of neutrinos produced in the interior of stars, LIM will be sensitive to the cosmic neutrino background (as CMB studies are, but at very different redshifts). Moreover, the energy of the neutrinos involved in each probe also varies, which may inform about a potential energy dependence of the electromagnetic transition moments~\cite{Frere:1996gb}. These synergies are very timely, since an enhanced magnetic moment may explain the $\sim 3\sigma$ excess observed by XENON1T~\cite{Aprile:2020tmw}, but the values require are close to the limits found by Borexino and in tension with stellar cooling constraints.

Finally, LIM may provide additional information about the cosmic neutrino background beyond the effect of $\sum m_\nu$ in the growth of perturbations: combining the information about $\sum m_\nu$ with the frequency of the photons produced in the decay, LIM might be the only cosmological probe sensitive to individual neutrino masses and their hierarchy~\cite{Archidiacono_nuhierarchy}.

The complementarity between different probes of neutrino decays will also help as a cross-check for eventual caveats or systematic uncertainties in the measurements. In the case of LIM experiments, these are the same as for the search for radiative dark matter decays, which are discussed in Ref.~\cite{Bernal_LIMDM}. In summary, astrophysical uncertainties are already accounted for in our analysis, there are efficient strategies to deal with known astrophysical line interlopers~\cite{Oh2003,Wang2006,Liu2011,Breysse_foregrounds, Lidz_interlopers, Sun_foregrounds, Cheng_foregroundsAP, Cheng_deconfInterlopers, Gong_interlopers}, and galactic foregrounds  are expected to be under control at the frequencies of interest. Moreover, the neutrino decay contribution to the LIM power spectrum and VID is very characteristic, and the combination of both summary statistics will not only improve the sensitivity but also the robustness of the measurement.\footnote{Here we consider the LIM power spectrum and VID constraints separately, but they could be combined if the covariance between them is available~\cite{Ihle_VID-PS}.} Finally, we have assumed that the neutrino decay line is a delta function, and neglected any widening due to the neutrino velocity distributions. While this is a good approximation for the regime of interest at this stage, it is also possible to model the neutrino decay emissivity with a generic momentum distribution~\cite{Aalberts:2018obr}; this will allow to adapt our analysis to neutrino production models that alter their momentum distribution~\cite{Merle:2015oja}.

The neutrino decay contribution might be confused with other exotic radiation injection such as dark matter decay. However, the shape of the neutrino power spectrum and density PDF is different. Moreover, while the contribution from dark matter decays will appear in LIM cross-correlations with galaxy clustering~\cite{Creque-Sarbinowski_LIMDM} and lensing~\cite{Shirasaki_limlensingdm}, the contribution from neutrino decays will barely do, since galaxy surveys do not trace the neutrino density field.

In this letter we have proposed the use of LIM for the detection of a possible radiative decay of the cosmic neutrino background, focusing on its contribution to the LIM power spectrum and VID. We have also provided a first parametric fit of the neutrino density PDF using N-body simulations and symbolic regression, that was required to compute the contribution to the VID. Our results show that LIM have the potential to achieve sensitivities competitive to current limits, improving other cosmological probes by several orders of magnitude. The complementarity of LIM and other existing probes of neutrino decays opens exciting synergies, as well as checks for systematics, that will lead the way to new studies of neutrino properties.

\textit{Acknowledgments---} 
JLB is supported by the Allan C. and Dorothy H. Davis Fellowship. AC acknowledges support from the the Israel Science Foundation (Grant No. 1302/19), the US-Israeli BSF (grant 2018236) and the German Israeli GIF (grant I-2524-303.7). AC acknowledges hospitality of the Max Planck Institute of Physics in Munich.  FVN acknowledges funding from the WFIRST program through NNG26PJ30C and NNN12AA01C. This work was supported at Johns Hopkins by NSF Grant No.\ 1818899 and the Simons Foundation.

\bibliography{Refs}

\providecommand{\href}[2]{#2}\begingroup\raggedright\begin{thebibliography}{10}

\bibitem{Yoshi75}
V.~Cung and M.~Yoshimura, ``Electromagnetic interaction of neutrinos in gauge
  theories of weak interactions,''
  \href{http://dx.doi.org/10.1007/BF02734528}{{\em La Rivista del Nuovo
  Cimento} {\bfseries 29} no.~4, (Oct., 1975) 557--564}.

\bibitem{Rosenberg:1962pp}
L.~Rosenberg, ``{Electromagnetic interactions of neutrinos},''
  \href{http://dx.doi.org/10.1103/PhysRev.129.2786}{{\em Phys. Rev.} {\bfseries
  129} (1963) 2786--2788}.

\bibitem{Mertens:2018vuu}
{\bfseries KATRIN} Collaboration, S.~Mertens {\em et~al.}, ``{A novel detector
  system for KATRIN to search for keV-scale sterile neutrinos},''
  \href{http://dx.doi.org/10.1088/1361-6471/ab12fe}{{\em J. Phys. G} {\bfseries
  46} no.~6, (2019) 065203}, \href{http://arxiv.org/abs/1810.06711}{{\ttfamily
  arXiv:1810.06711 [physics.ins-det]}}.

\bibitem{KamLAND-Zen:2016pfg}
{\bfseries KamLAND-Zen} Collaboration, A.~Gando {\em et~al.}, ``{Search for
  Majorana Neutrinos near the Inverted Mass Hierarchy Region with
  KamLAND-Zen},'' \href{http://dx.doi.org/10.1103/PhysRevLett.117.082503}{{\em
  Phys. Rev. Lett.} {\bfseries 117} no.~8, (2016) 082503},
  \href{http://arxiv.org/abs/1605.02889}{{\ttfamily arXiv:1605.02889
  [hep-ex]}}. [Addendum: Phys.Rev.Lett. 117, 109903 (2016)].

\bibitem{Albert:2014awa}
{\bfseries EXO-200} Collaboration, J.~B. Albert {\em et~al.}, ``{Search for
  Majorana neutrinos with the first two years of EXO-200 data},''
  \href{http://dx.doi.org/10.1038/nature13432}{{\em Nature} {\bfseries 510}
  (2014) 229--234}, \href{http://arxiv.org/abs/1402.6956}{{\ttfamily
  arXiv:1402.6956 [nucl-ex]}}.

\bibitem{Brunner:2017iql}
T.~Brunner and L.~Winslow, ``{Searching for $0\nu\beta\beta$ decay in
  $^{136}$Xe -- towards the tonne-scale and beyond},''
  \href{http://dx.doi.org/10.1080/10619127.2017.1315286}{{\em Nucl. Phys. News}
  {\bfseries 27} no.~3, (2017) 14--19},
  \href{http://arxiv.org/abs/1704.01528}{{\ttfamily arXiv:1704.01528
  [hep-ex]}}.

\bibitem{Beda:2009kx}
A.~Beda, E.~Demidova, A.~Starostin, V.~Brudanin, V.~Egorov, D.~Medvedev, {\em
  et~al.}, ``{GEMMA experiment: Three years of the search for the neutrino
  magnetic moment},'' \href{http://dx.doi.org/10.1134/S1547477110060063}{{\em
  Phys. Part. Nucl. Lett.} {\bfseries 7} (2010) 406--409},
  \href{http://arxiv.org/abs/0906.1926}{{\ttfamily arXiv:0906.1926 [hep-ex]}}.

\bibitem{Deniz:2009mu}
{\bfseries TEXONO} Collaboration, M.~Deniz {\em et~al.}, ``{Measurement of
  Nu(e)-bar -Electron Scattering Cross-Section with a CsI(Tl) Scintillating
  Crystal Array at the Kuo-Sheng Nuclear Power Reactor},''
  \href{http://dx.doi.org/10.1103/PhysRevD.81.072001}{{\em Phys. Rev. D}
  {\bfseries 81} (2010) 072001},
  \href{http://arxiv.org/abs/0911.1597}{{\ttfamily arXiv:0911.1597 [hep-ex]}}.

\bibitem{Abe:2017vif}
{\bfseries T2K} Collaboration, K.~Abe {\em et~al.}, ``{Measurement of neutrino
  and antineutrino oscillations by the T2K experiment including a new
  additional sample of $\nu_e$ interactions at the far detector},''
  \href{http://dx.doi.org/10.1103/PhysRevD.96.092006}{{\em Phys. Rev. D}
  {\bfseries 96} no.~9, (2017) 092006},
  \href{http://arxiv.org/abs/1707.01048}{{\ttfamily arXiv:1707.01048
  [hep-ex]}}. [Erratum: Phys.Rev.D 98, 019902 (2018)].

\bibitem{Adamson:2017gxd}
{\bfseries NOvA} Collaboration, P.~Adamson {\em et~al.}, ``{Constraints on
  Oscillation Parameters from $\nu_e$ Appearance and $\nu_\mu$ Disappearance in
  NOvA},'' \href{http://dx.doi.org/10.1103/PhysRevLett.118.231801}{{\em Phys.
  Rev. Lett.} {\bfseries 118} no.~23, (2017) 231801},
  \href{http://arxiv.org/abs/1703.03328}{{\ttfamily arXiv:1703.03328
  [hep-ex]}}.

\bibitem{Acciarri:2016crz}
{\bfseries DUNE} Collaboration, R.~Acciarri {\em et~al.}, ``{Long-Baseline
  Neutrino Facility (LBNF) and Deep Underground Neutrino Experiment (DUNE)}:
  {Conceptual Design Report, Volume 1: The LBNF and DUNE Projects},''
  \href{http://arxiv.org/abs/1601.05471}{{\ttfamily arXiv:1601.05471
  [physics.ins-det]}}.

\bibitem{Abe:2015zbg}
{\bfseries Hyper-Kamiokande Proto-} Collaboration, K.~Abe {\em et~al.},
  ``{Physics potential of a long-baseline neutrino oscillation experiment using
  a J-PARC neutrino beam and Hyper-Kamiokande},''
  \href{http://dx.doi.org/10.1093/ptep/ptv061}{{\em PTEP} {\bfseries 2015}
  (2015) 053C02}, \href{http://arxiv.org/abs/1502.05199}{{\ttfamily
  arXiv:1502.05199 [hep-ex]}}.

\bibitem{Aartsen:2017bap}
{\bfseries IceCube} Collaboration, M.~G. Aartsen {\em et~al.}, ``{Search for
  sterile neutrino mixing using three years of IceCube DeepCore data},''
  \href{http://dx.doi.org/10.1103/PhysRevD.95.112002}{{\em Phys. Rev. D}
  {\bfseries 95} no.~11, (2017) 112002},
  \href{http://arxiv.org/abs/1702.05160}{{\ttfamily arXiv:1702.05160
  [hep-ex]}}.

\bibitem{Raffelt:1996wa}
G.~Raffelt, {\em {Stars as laboratories for fundamental physics}: {The
  astrophysics of neutrinos, axions, and other weakly interacting particles}}.
\newblock 5, 1996.

\bibitem{Pontecorvo:1967fh}
B.~Pontecorvo, ``{Neutrino Experiments and the Problem of Conservation of
  Leptonic Charge},'' {\em Sov. Phys. JETP} {\bfseries 26} (1968) 984--988.

\bibitem{Gribov:1968kq}
V.~Gribov and B.~Pontecorvo, ``{Neutrino astronomy and lepton charge},''
  \href{http://dx.doi.org/10.1016/0370-2693(69)90525-5}{{\em Phys. Lett. B}
  {\bfseries 28} (1969) 493}.

\bibitem{Langacker:1986jv}
P.~Langacker, S.~Petcov, G.~Steigman, and S.~Toshev, ``{On the
  Mikheev-Smirnov-Wolfenstein (MSW) Mechanism of Amplification of Neutrino
  Oscillations in Matter},''
  \href{http://dx.doi.org/10.1016/0550-3213(87)90699-7}{{\em Nucl. Phys. B}
  {\bfseries 282} (1987) 589--609}.

\bibitem{Bilenky:1980cx}
S.~M. Bilenky, J.~Hosek, and S.~Petcov, ``{On Oscillations of Neutrinos with
  Dirac and Majorana Masses},''
  \href{http://dx.doi.org/10.1016/0370-2693(80)90927-2}{{\em Phys. Lett. B}
  {\bfseries 94} (1980) 495--498}.

\bibitem{Akhmedov:1999uz}
E.~K. Akhmedov, ``{Neutrino physics},'' in {\em {ICTP Summer School in Particle
  Physics}}, pp.~103--164.
\newblock 6, 1999.
\newblock \href{http://arxiv.org/abs/hep-ph/0001264}{{\ttfamily
  arXiv:hep-ph/0001264}}.

\bibitem{GonzalezGarcia:2002dz}
M.~C. Gonzalez-Garcia and Y.~Nir, ``{Neutrino Masses and Mixing: Evidence and
  Implications},'' \href{http://dx.doi.org/10.1103/RevModPhys.75.345}{{\em Rev.
  Mod. Phys.} {\bfseries 75} (2003) 345--402},
  \href{http://arxiv.org/abs/hep-ph/0202058}{{\ttfamily arXiv:hep-ph/0202058}}.

\bibitem{GonzalezGarcia:2007ib}
M.~C. Gonzalez-Garcia and M.~Maltoni, ``{Phenomenology with Massive
  Neutrinos},'' \href{http://dx.doi.org/10.1016/j.physrep.2007.12.004}{{\em
  Phys. Rept.} {\bfseries 460} (2008) 1--129},
  \href{http://arxiv.org/abs/0704.1800}{{\ttfamily arXiv:0704.1800 [hep-ph]}}.

\bibitem{1992pmn..book.....B}
F.~{Boehm} and P.~{Vogel}, {\em {Physics of Massive Neutrinos}}.
\newblock 1992.

\bibitem{Pal:1981rm}
P.~B. Pal and L.~Wolfenstein, ``{Radiative Decays of Massive Neutrinos},''
  \href{http://dx.doi.org/10.1103/PhysRevD.25.766}{{\em Phys. Rev. D}
  {\bfseries 25} (1982) 766}.

\bibitem{Petcov:1976ff}
S.~T. Petcov, ``{The Processes $\mu \rightarrow e + \gamma, \mu \rightarrow e +
  \overline{e}, \nu' \rightarrow \nu + \gamma$ in the Weinberg-Salam Model with
  Neutrino Mixing},'' {\em Sov. J. Nucl. Phys.} {\bfseries 25} (1977) 340.
  [Erratum: Sov.J.Nucl.Phys. 25, 698 (1977), Erratum: Yad.Fiz. 25, 1336
  (1977)].

\bibitem{Fujikawa:1980yx}
K.~Fujikawa and R.~Shrock, ``{The Magnetic Moment of a Massive Neutrino and
  Neutrino Spin Rotation},''
  \href{http://dx.doi.org/10.1103/PhysRevLett.45.963}{{\em Phys. Rev. Lett.}
  {\bfseries 45} (1980) 963}.

\bibitem{Bell:2006wi}
N.~F. Bell, M.~Gorchtein, M.~J. Ramsey-Musolf, P.~Vogel, and P.~Wang, ``{Model
  independent bounds on magnetic moments of Majorana neutrinos},''
  \href{http://dx.doi.org/10.1016/j.physletb.2006.09.055}{{\em Phys. Lett. B}
  {\bfseries 642} (2006) 377--383},
  \href{http://arxiv.org/abs/hep-ph/0606248}{{\ttfamily arXiv:hep-ph/0606248}}.

\bibitem{Davidson:2005cs}
S.~Davidson, M.~Gorbahn, and A.~Santamaria, ``{From transition magnetic moments
  to majorana neutrino masses},''
  \href{http://dx.doi.org/10.1016/j.physletb.2005.08.086}{{\em Phys. Lett. B}
  {\bfseries 626} (2005) 151--160},
  \href{http://arxiv.org/abs/hep-ph/0506085}{{\ttfamily arXiv:hep-ph/0506085}}.

\bibitem{Georgi:1990se}
H.~Georgi and L.~Randall, ``{Charge Conjugation and Neutrino Magnetic
  Moments},'' \href{http://dx.doi.org/10.1016/0370-2693(90)90055-B}{{\em Phys.
  Lett. B} {\bfseries 244} (1990) 196--202}.

\bibitem{Lee:1977tib}
B.~W. Lee and R.~E. Shrock, ``{Natural Suppression of Symmetry Violation in
  Gauge Theories: Muon - Lepton and Electron Lepton Number Nonconservation},''
  \href{http://dx.doi.org/10.1103/PhysRevD.16.1444}{{\em Phys. Rev. D}
  {\bfseries 16} (1977) 1444}.

\bibitem{Shrock:1974nd}
R.~Shrock, ``{Decay l0 ---\ensuremath{>} nu(lepton) gamma in gauge theories of
  weak and electromagnetic interactions},''
  \href{http://dx.doi.org/10.1103/PhysRevD.9.743}{{\em Phys. Rev. D} {\bfseries
  9} (1974) 743--748}.

\bibitem{Lindner:2017uvt}
M.~Lindner, B.~Radov\v{c}i\'c, and J.~Welter, ``{Revisiting Large Neutrino
  Magnetic Moments},'' \href{http://dx.doi.org/10.1007/JHEP07(2017)139}{{\em
  JHEP} {\bfseries 07} (2017) 139},
  \href{http://arxiv.org/abs/1706.02555}{{\ttfamily arXiv:1706.02555
  [hep-ph]}}.

\bibitem{Lindner_2017}
M.~Lindner, B.~Radovčić, and J.~Welter, ``Revisiting large neutrino magnetic
  moments,'' \href{http://dx.doi.org/10.1007/jhep07(2017)139}{{\em Journal of
  High Energy Physics} {\bfseries 2017} no.~7, (Jul, 2017) }.
  \url{http://dx.doi.org/10.1007/JHEP07(2017)139}.

\bibitem{Miranda:2020kwy}
O.~G. Miranda, D.~K. Papoulias, M.~T\'ortola, and J.~W.~F. Valle, ``{XENON1T
  signal from transition neutrino magnetic moments},''
  \href{http://dx.doi.org/10.1016/j.physletb.2020.135685}{{\em Phys. Lett. B}
  {\bfseries 808} (2020) 135685},
  \href{http://arxiv.org/abs/2007.01765}{{\ttfamily arXiv:2007.01765
  [hep-ph]}}.

\bibitem{Babu:2020ivd}
K.~S. Babu, S.~Jana, and M.~Lindner, ``Large neutrino magnetic moments in the
  light of recent experiments,''
  \href{http://dx.doi.org/10.1007/jhep10(2020)040}{{\em Journal of High Energy
  Physics} {\bfseries 2020} no.~10, (Oct, 2020) }.
  \url{http://dx.doi.org/10.1007/JHEP10(2020)040}.

\bibitem{Bahcall:1989ks}
J.~N. Bahcall, {\em {NEUTRINO ASTROPHYSICS}}.
\newblock 1989.

\bibitem{Aprile:2020tmw}
{\bfseries XENON} Collaboration, E.~Aprile {\em et~al.}, ``{Excess electronic
  recoil events in XENON1T},''
  \href{http://dx.doi.org/10.1103/PhysRevD.102.072004}{{\em Phys. Rev. D}
  {\bfseries 102} no.~7, (2020) 072004},
  \href{http://arxiv.org/abs/2006.09721}{{\ttfamily arXiv:2006.09721
  [hep-ex]}}.

\bibitem{Giunti:2014ixa}
C.~Giunti and A.~Studenikin, ``{Neutrino electromagnetic interactions: a window
  to new physics},'' \href{http://dx.doi.org/10.1103/RevModPhys.87.531}{{\em
  Rev. Mod. Phys.} {\bfseries 87} (2015) 531},
  \href{http://arxiv.org/abs/1403.6344}{{\ttfamily arXiv:1403.6344 [hep-ph]}}.

\bibitem{PhysRevD.39.3378}
P.~Vogel and J.~Engel, ``Neutrino electromagnetic form factors,''
  \href{http://dx.doi.org/10.1103/PhysRevD.39.3378}{{\em Phys. Rev. D}
  {\bfseries 39} (Jun, 1989) 3378--3383}.
  \url{https://link.aps.org/doi/10.1103/PhysRevD.39.3378}.

\bibitem{PhysRevLett.64.2856}
G.~G. Raffelt, ``New bound on neutrino dipole moments from globular-cluster
  stars,'' \href{http://dx.doi.org/10.1103/PhysRevLett.64.2856}{{\em Phys. Rev.
  Lett.} {\bfseries 64} (Jun, 1990) 2856--2858}.
  \url{https://link.aps.org/doi/10.1103/PhysRevLett.64.2856}.

\bibitem{1992A&A...264..536R}
G.~{Raffelt} and A.~{Weiss}, ``{Non-standard neutrino interactions and the
  evolution of red giants},'' {\em \aap} {\bfseries 264} no.~2, (Oct., 1992)
  536--546.

\bibitem{Viaux:2013hca}
N.~Viaux, M.~Catelan, P.~B. Stetson, G.~Raffelt, J.~Redondo, A.~A.~R. Valcarce,
  and A.~Weiss, ``{Particle-physics constraints from the globular cluster M5:
  Neutrino Dipole Moments},''
  \href{http://dx.doi.org/10.1051/0004-6361/201322004}{{\em Astron. Astrophys.}
  {\bfseries 558} (2013) A12}, \href{http://arxiv.org/abs/1308.4627}{{\ttfamily
  arXiv:1308.4627 [astro-ph.SR]}}.

\bibitem{Mirizzi:2007jd}
A.~Mirizzi, D.~Montanino, and P.~D. Serpico, ``{Revisiting cosmological bounds
  on radiative neutrino lifetime},''
  \href{http://dx.doi.org/10.1103/PhysRevD.76.053007}{{\em Phys. Rev. D}
  {\bfseries 76} (2007) 053007},
  \href{http://arxiv.org/abs/0705.4667}{{\ttfamily arXiv:0705.4667 [hep-ph]}}.

\bibitem{Aalberts:2018obr}
J.~L. Aalberts {\em et~al.}, ``{Precision constraints on radiative neutrino
  decay with CMB spectral distortion},''
  \href{http://dx.doi.org/10.1103/PhysRevD.98.023001}{{\em Phys. Rev. D}
  {\bfseries 98} (2018) 023001},
  \href{http://arxiv.org/abs/1803.00588}{{\ttfamily arXiv:1803.00588
  [astro-ph.CO]}}.

\bibitem{Kovetz:2017agg}
E.~D. Kovetz {\em et~al.}, ``{Line-Intensity Mapping: 2017 Status Report},''
  \href{http://arxiv.org/abs/1709.09066}{{\ttfamily arXiv:1709.09066
  [astro-ph.CO]}}.

\bibitem{Kovetz:2019uss}
E.~D. Kovetz {\em et~al.}, ``{Astrophysics and Cosmology with Line-Intensity
  Mapping},'' \href{http://arxiv.org/abs/1903.04496}{{\ttfamily
  arXiv:1903.04496 [astro-ph.CO]}}.

\bibitem{Creque-Sarbinowski_LIMDM}
C.~{Creque-Sarbinowski} and M.~{Kamionkowski}, ``{Searching for decaying and
  annihilating dark matter with line intensity mapping},''
  \href{http://dx.doi.org/10.1103/PhysRevD.98.063524}{{\em \prd} {\bfseries 98}
  no.~6, (Sep, 2018) 063524}, \href{http://arxiv.org/abs/1806.11119}{{\ttfamily
  arXiv:1806.11119 [astro-ph.CO]}}.

\bibitem{Bernal_LIMDM}
J.~L. Bernal, A.~Caputo, and M.~Kamionkowski, ``{Strategies to Detect
  Dark-Matter Decays with Line-Intensity Mapping},''
  \href{http://arxiv.org/abs/2012.00771}{{\ttfamily arXiv:2012.00771
  [astro-ph.CO]}}.

\bibitem{Mangano:2006ar}
G.~Mangano, G.~Miele, S.~Pastor, T.~Pinto, O.~Pisanti, and P.~D. Serpico,
  ``{Effects of non-standard neutrino-electron interactions on relic neutrino
  decoupling},'' \href{http://dx.doi.org/10.1016/j.nuclphysb.2006.09.002}{{\em
  Nucl. Phys. B} {\bfseries 756} (2006) 100--116},
  \href{http://arxiv.org/abs/hep-ph/0607267}{{\ttfamily arXiv:hep-ph/0607267}}.

\bibitem{Lidz_2011}
A.~{Lidz}, S.~R. {Furlanetto}, S.~P. {Oh}, J.~{Aguirre}, T.-C. {Chang},
  O.~{Dor{\'e}}, and J.~R. {Pritchard}, ``{Intensity Mapping with Carbon
  Monoxide Emission Lines and the Redshifted 21 cm Line},''
  \href{http://dx.doi.org/10.1088/0004-637X/741/2/70}{{\em \apj} {\bfseries
  741} (Nov., 2011) 70}, \href{http://arxiv.org/abs/1104.4800}{{\ttfamily
  arXiv:1104.4800}}.

\bibitem{Bernal_IM}
J.~L. {Bernal}, P.~C. {Breysse}, H.~{Gil-Mar{\'\i}n}, and E.~D. {Kovetz},
  ``{User's guide to extracting cosmological information from line-intensity
  maps},'' \href{http://dx.doi.org/10.1103/PhysRevD.100.123522}{{\em \prd}
  {\bfseries 100} no.~12, (Dec., 2019) 123522},
  \href{http://arxiv.org/abs/1907.10067}{{\ttfamily arXiv:1907.10067
  [astro-ph.CO]}}.

\bibitem{Lewis:1999bs}
A.~Lewis, A.~Challinor, and A.~Lasenby, ``{Efficient computation of CMB
  anisotropies in closed FRW models},''
  \href{http://dx.doi.org/10.1086/309179}{{\em Astrophys. J.} {\bfseries 538}
  (2000) 473--476}, \href{http://arxiv.org/abs/astro-ph/9911177}{{\ttfamily
  arXiv:astro-ph/9911177}}.

\bibitem{Villaescusa-Navarro:2019bje}
F.~Villaescusa-Navarro {\em et~al.}, ``{The Quijote simulations},''
  \href{http://dx.doi.org/10.3847/1538-4365/ab9d82}{{\em Astrophys. J. Suppl.}
  {\bfseries 250} no.~1, (2020) 2},
  \href{http://arxiv.org/abs/1909.05273}{{\ttfamily arXiv:1909.05273
  [astro-ph.CO]}}.

\bibitem{Oh2003}
S.~P. {Oh} and K.~J. {Mack}, ``{Foregrounds for 21-cm observations of neutral
  gas at high redshift},''
  \href{http://dx.doi.org/10.1111/j.1365-2966.2003.07133.x}{{\em \mnras}
  {\bfseries 346} no.~3, (Dec, 2003) 871--877},
  \href{http://arxiv.org/abs/astro-ph/0302099}{{\ttfamily
  arXiv:astro-ph/0302099 [astro-ph]}}.

\bibitem{Wang2006}
X.~{Wang}, M.~{Tegmark}, M.~G. {Santos}, and L.~{Knox}, ``{21 cm Tomography
  with Foregrounds},'' \href{http://dx.doi.org/10.1086/506597}{{\em \apj}
  {\bfseries 650} no.~2, (Oct, 2006) 529--537},
  \href{http://arxiv.org/abs/astro-ph/0501081}{{\ttfamily
  arXiv:astro-ph/0501081 [astro-ph]}}.

\bibitem{Liu2011}
A.~{Liu} and M.~{Tegmark}, ``{A method for 21 cm power spectrum estimation in
  the presence of foregrounds},''
  \href{http://dx.doi.org/10.1103/PhysRevD.83.103006}{{\em \prd} {\bfseries 83}
  no.~10, (May, 2011) 103006}, \href{http://arxiv.org/abs/1103.0281}{{\ttfamily
  arXiv:1103.0281 [astro-ph.CO]}}.

\bibitem{Breysse_foregrounds}
P.~C. {Breysse}, E.~D. {Kovetz}, and M.~{Kamionkowski}, ``{Masking line
  foregrounds in intensity-mapping surveys},''
  \href{http://dx.doi.org/10.1093/mnras/stv1476}{{\em \mnras} {\bfseries 452}
  no.~4, (Oct, 2015) 3408--3418},
  \href{http://arxiv.org/abs/1503.05202}{{\ttfamily arXiv:1503.05202
  [astro-ph.CO]}}.

\bibitem{Lidz_interlopers}
A.~{Lidz} and J.~{Taylor}, ``{On Removing Interloper Contamination from
  Intensity Mapping Power Spectrum Measurements},''
  \href{http://dx.doi.org/10.3847/0004-637X/825/2/143}{{\em \apj} {\bfseries
  825} no.~2, (July, 2016) 143},
  \href{http://arxiv.org/abs/1604.05737}{{\ttfamily arXiv:1604.05737
  [astro-ph.CO]}}.

\bibitem{Sun_foregrounds}
G.~{Sun}, L.~{Moncelsi}, M.~P. {Viero}, M.~B. {Silva}, J.~{Bock}, C.~M.
  {Bradford}, {\em et~al.}, ``{A Foreground Masking Strategy for [C II]
  Intensity Mapping Experiments Using Galaxies Selected by Stellar Mass and
  Redshift},'' \href{http://dx.doi.org/10.3847/1538-4357/aab3e3}{{\em \apj}
  {\bfseries 856} no.~2, (Apr, 2018) 107},
  \href{http://arxiv.org/abs/1610.10095}{{\ttfamily arXiv:1610.10095
  [astro-ph.GA]}}.

\bibitem{Cheng_foregroundsAP}
Y.-T. {Cheng}, T.-C. {Chang}, J.~{Bock}, C.~M. {Bradford}, and A.~{Cooray},
  ``{Spectral Line De-confusion in an Intensity Mapping Survey},''
  \href{http://dx.doi.org/10.3847/0004-637X/832/2/165}{{\em \apj} {\bfseries
  832} no.~2, (Dec, 2016) 165},
  \href{http://arxiv.org/abs/1604.07833}{{\ttfamily arXiv:1604.07833
  [astro-ph.CO]}}.

\bibitem{Cheng_deconfInterlopers}
Y.-T. {Cheng}, T.-C. {Chang}, and J.~J. {Bock}, ``{Phase-space Spectral Line
  Deconfusion in Intensity Mapping},''
  \href{http://dx.doi.org/10.3847/1538-4357/abb023}{{\em \apj} {\bfseries 901}
  no.~2, (Oct., 2020) 142}, \href{http://arxiv.org/abs/2005.05341}{{\ttfamily
  arXiv:2005.05341 [astro-ph.CO]}}.

\bibitem{Gong_interlopers}
Y.~{Gong}, X.~{Chen}, and A.~{Cooray}, ``{Cosmological Constraints from Line
  Intensity Mapping with Interlopers},''
  \href{http://dx.doi.org/10.3847/1538-4357/ab87a0}{{\em \apj} {\bfseries 894}
  no.~2, (May, 2020) 152}, \href{http://arxiv.org/abs/2001.10792}{{\ttfamily
  arXiv:2001.10792 [astro-ph.CO]}}.

\bibitem{Cleary_COMAP}
K.~{Cleary}, M.-A. {Bigot-Sazy}, D.~{Chung}, {\em et~al.}, ``{The CO Mapping
  Array Pathfinder (COMAP)},'' {\em AAS Meeting Abstracts \#227} {\bfseries
  227} (Jan., 2016) 426.06.

\bibitem{CCAT-prime}
S.~K. {Choi} {\em et~al.}, ``{Sensitivity of the Prime-Cam Instrument on the
  CCAT-Prime Telescope},''
  \href{http://dx.doi.org/10.1007/s10909-020-02428-z}{{\em Journal of Low
  Temperature Physics} {\bfseries 199} no.~3-4, (Mar., 2020) 1089--1097},
  \href{http://arxiv.org/abs/1908.10451}{{\ttfamily arXiv:1908.10451
  [astro-ph.IM]}}.

\bibitem{ATLAST}
P.~D. {Klaassen}, T.~K. {Mroczkowski}, C.~{Cicone}, E.~{Hatziminaoglou},
  S.~{Sartori}, C.~{De Breuck}, {\em et~al.},
  \href{http://dx.doi.org/10.1117/12.2561315}{``{The Atacama Large Aperture
  Submillimeter Telescope (AtLAST)},''} in {\em Society of Photo-Optical
  Instrumentation Engineers (SPIE) Conference Series}, vol.~11445 of {\em
  Society of Photo-Optical Instrumentation Engineers (SPIE) Conference Series},
  p.~114452F.
\newblock Dec., 2020.
\newblock \href{http://arxiv.org/abs/2011.07974}{{\ttfamily arXiv:2011.07974
  [astro-ph.IM]}}.

\bibitem{Li_CO_16}
T.~Y. {Li}, R.~H. {Wechsler}, K.~{Devaraj}, and S.~E. {Church}, ``{Connecting
  CO Intensity Mapping to Molecular Gas and Star Formation in the Epoch of
  Galaxy Assembly},'' \href{http://dx.doi.org/10.3847/0004-637X/817/2/169}{{\em
  \apj} {\bfseries 817} (Feb., 2016) 169},
  \href{http://arxiv.org/abs/1503.08833}{{\ttfamily arXiv:1503.08833
  [astro-ph.CO]}}.

\bibitem{Silva_CII}
M.~{Silva}, M.~G. {Santos}, A.~{Cooray}, and Y.~{Gong}, ``{Prospects for
  Detecting C II Emission during the Epoch of Reionization},''
  \href{http://dx.doi.org/10.1088/0004-637X/806/2/209}{{\em \apj} {\bfseries
  806} no.~2, (Jun, 2015) 209},
  \href{http://arxiv.org/abs/1410.4808}{{\ttfamily arXiv:1410.4808
  [astro-ph.GA]}}.

\bibitem{Breysse_VID}
P.~C. {Breysse}, E.~D. {Kovetz}, P.~S. {Behroozi}, L.~{Dai}, and
  M.~{Kamionkowski}, ``{Insights from probability distribution functions of
  intensity maps},'' \href{http://dx.doi.org/10.1093/mnras/stx203}{{\em \mnras}
  {\bfseries 467} (May, 2017) 2996--3010},
  \href{http://arxiv.org/abs/1609.01728}{{\ttfamily arXiv:1609.01728
  [astro-ph.CO]}}.

\bibitem{Planck18_parameters}
{Planck Collaboration}, N.~Aghanim, Y.~Akrami, M.~Ashdown, J.~Aumont,
  C.~Baccigalupi, {\em et~al.}, ``{Planck 2018 results: VI. Cosmological
  parameters},'' \href{http://dx.doi.org/10.1051/0004-6361/201833910}{{\em
  Astronomy and Astrophysics} {\bfseries 641} (Sep, 2020) A6},
  \href{http://arxiv.org/abs/1807.06209}{{\ttfamily arXiv:1807.06209}}.
  \url{https://doi.org/10.1051/0004-6361/201833910}.

\bibitem{Fisher:1935}
R.~A. Fisher, ``{The Fiducial Argument in Statistical Inference},''
\href{http://dx.doi.org/10.1111/j.1469-1809.1935.tb02120.x}{{\em Annals Eugen.}
  {\bfseries 6} (1935) 391--398}.

\bibitem{Jungman:1995av}
G.~Jungman, M.~Kamionkowski, A.~Kosowsky, and D.~N. Spergel, ``{Weighing the
  universe with the cosmic microwave background},''
  \href{http://dx.doi.org/10.1103/PhysRevLett.76.1007}{{\em Phys. Rev. Lett.}
  {\bfseries 76} (1996) 1007--1010},
  \href{http://arxiv.org/abs/astro-ph/9507080}{{\ttfamily
  arXiv:astro-ph/9507080}}.

\bibitem{Jungman:1995bz}
G.~Jungman, M.~Kamionkowski, A.~Kosowsky, and D.~N. Spergel, ``{Cosmological
  parameter determination with microwave background maps},''
  \href{http://dx.doi.org/10.1103/PhysRevD.54.1332}{{\em Phys. Rev. D}
  {\bfseries 54} (1996) 1332--1344},
  \href{http://arxiv.org/abs/astro-ph/9512139}{{\ttfamily
  arXiv:astro-ph/9512139}}.

\bibitem{Tegmark_fisher97}
M.~{Tegmark}, A.~N. {Taylor}, and A.~F. {Heavens}, ``{Karhunen-Lo{\`e}ve
  Eigenvalue Problems in Cosmology: How Should We Tackle Large Data Sets?},''
  \href{http://dx.doi.org/10.1086/303939}{{\em \apj} {\bfseries 480} (May,
  1997) 22--35}, \href{http://arxiv.org/abs/astro-ph/9603021}{{\ttfamily
  astro-ph/9603021}}.

\bibitem{Esteban:2018azc}
I.~Esteban, M.~Gonzalez-Garcia, A.~Hernandez-Cabezudo, M.~Maltoni, and
  T.~Schwetz, ``{Global analysis of three-flavour neutrino oscillations:
  synergies and tensions in the determination of $\theta_{23}$, $\delta_{CP}$,
  and the mass ordering},''
  \href{http://dx.doi.org/10.1007/JHEP01(2019)106}{{\em JHEP} {\bfseries 01}
  (2019) 106}, \href{http://arxiv.org/abs/1811.05487}{{\ttfamily
  arXiv:1811.05487 [hep-ph]}}.

\bibitem{Brdar:2020quo}
V.~Brdar, A.~Greljo, J.~Kopp, and T.~Opferkuch, ``{The Neutrino Magnetic Moment
  Portal: Cosmology, Astrophysics, and Direct Detection},''
  \href{http://dx.doi.org/10.1088/1475-7516/2021/01/039}{{\em JCAP} {\bfseries
  01} (2021) 039}, \href{http://arxiv.org/abs/2007.15563}{{\ttfamily
  arXiv:2007.15563 [hep-ph]}}.

\bibitem{Borexino:2017fbd}
{\bfseries Borexino} Collaboration, M.~Agostini {\em et~al.}, ``{Limiting
  neutrino magnetic moments with Borexino Phase-II solar neutrino data},''
  \href{http://dx.doi.org/10.1103/PhysRevD.96.091103}{{\em Phys. Rev. D}
  {\bfseries 96} no.~9, (2017) 091103},
  \href{http://arxiv.org/abs/1707.09355}{{\ttfamily arXiv:1707.09355
  [hep-ex]}}.

\bibitem{Frere:1996gb}
J.~M. Frere, R.~B. Nevzorov, and M.~I. Vysotsky, ``{Stimulated neutrino
  conversion and bounds on neutrino magnetic moments},''
  \href{http://dx.doi.org/10.1016/S0370-2693(96)01667-X}{{\em Phys. Lett. B}
  {\bfseries 394} (1997) 127--131},
  \href{http://arxiv.org/abs/hep-ph/9608266}{{\ttfamily arXiv:hep-ph/9608266}}.

\bibitem{Archidiacono_nuhierarchy}
M.~{Archidiacono}, S.~{Hannestad}, and J.~{Lesgourgues}, ``{What will it take
  to measure individual neutrino mass states using cosmology?},''
  \href{http://dx.doi.org/10.1088/1475-7516/2020/09/021}{{\em \jcap} {\bfseries
  2020} no.~9, (Sept., 2020) 021},
  \href{http://arxiv.org/abs/2003.03354}{{\ttfamily arXiv:2003.03354
  [astro-ph.CO]}}.

\bibitem{Ihle_VID-PS}
H.~T. {Ihle}, D.~{Chung}, G.~{Stein}, M.~{Alvarez}, J.~R. {Bond}, P.~C.
  {Breysse}, {\em et~al.}, ``{Joint Power Spectrum and Voxel Intensity
  Distribution Forecast on the CO Luminosity Function with COMAP},''
  \href{http://dx.doi.org/10.3847/1538-4357/aaf4bc}{{\em \apj} {\bfseries 871}
  (Jan, 2019) 75}, \href{http://arxiv.org/abs/1808.07487}{{\ttfamily
  arXiv:1808.07487 [astro-ph.CO]}}.

\bibitem{Merle:2015oja}
A.~Merle and M.~Totzauer, ``{keV Sterile Neutrino Dark Matter from Singlet
  Scalar Decays: Basic Concepts and Subtle Features},''
  \href{http://dx.doi.org/10.1088/1475-7516/2015/06/011}{{\em JCAP} {\bfseries
  06} (2015) 011}, \href{http://arxiv.org/abs/1502.01011}{{\ttfamily
  arXiv:1502.01011 [hep-ph]}}.

\bibitem{Shirasaki_limlensingdm}
M.~{Shirasaki}, ``{Searching for eV-mass Axion-like Particles with Cross
  Correlations between Line Intensity and Weak Lensing Maps},'' {\em arXiv
  e-prints} (Jan., 2021) arXiv:2102.00580,
  \href{http://arxiv.org/abs/2102.00580}{{\ttfamily arXiv:2102.00580
  [astro-ph.CO]}}.

\end{thebibliography}\endgroup
\bibliographystyle{utcaps}
\bigskip

\end{document}